\documentclass[aps,prd,twocolumn,superscriptaddress]{revtex4}

\def\mn{m_{\nu}}
\def\nn{n_{\nu}}
\def\tn{T_{\nu}}
\def\en{{\cal E}_{\nu}}
\def\beq{\begin{equation}}
\def\eeq{\end{equation}}
\def\bea{\begin{eqnarray}}
\def\eea{\end{eqnarray}}
\def\a{{\cal A}}
\def\ev{~{\rm eV}}

\usepackage{graphicx}
\usepackage{dcolumn}
\usepackage{bm}
\begin{document}

\title{On the stability of Dark Energy with Mass-Varying Neutrinos}

\author{Niayesh Afshordi} \email{nafshordi@cfa.harvard.edu}
\affiliation{Institute for Theory and Computation,
Harvard-Smithsonian Center for Astrophysics, MS-51, 60 Garden
Street, Cambridge, MA 02138, USA}
\author{Matias Zaldarriaga}\email{mzaldarriaga@cfa.harvard.edu}
\affiliation{Institute for Theory and Computation,
Harvard-Smithsonian Center for Astrophysics, MS-51, 60 Garden
Street, Cambridge, MA 02138, USA}\affiliation{Jefferson Laboratory
of Physics, Harvard University, Cambridge, Massachusetts 02138, USA}
\author{Kazunori Kohri}\email{kkohri@cfa.harvard.edu}
\affiliation{Institute for Theory and Computation,
Harvard-Smithsonian Center for Astrophysics, MS-51, 60 Garden
Street, Cambridge, MA 02138, USA} \affiliation{Department of Earth
and Space Science, Graduate School of Science, Osaka University,
Osaka 560-0043, Japan}

\author{}
\affiliation{}


\date{\today}

\begin{abstract}
An interesting dynamical model for dark energy which does not
require extremely light scalar fields such as quintessence, and at
the same time explains the (near-) coincidence between the neutrino
and dark energy densities is the model of dark energy coupled to
mass varying neutrinos (MaVaNs). Despite the attractions of this
model, we show that, generically, this model contains a catastrophic
instability which occurs when neutrinos become non-relativistic. As
a result of this instability, as neutrinos become non-relativistic,
they condense into neutrino nuggets which redshift away similar to
cold dark matter, and thus cease to act as dark energy. Any stable
MaVaNs dark energy model is extremely contrived, and is virtually
indistinguishable from a cosmological constant.
\end{abstract}


\maketitle

\section{Introduction}

During the past decade, it has become clear that different
cosmological observations, such as the dimming of distant supernovae
Ia \cite{Riess,Perlmutter:1998np}, anisotropies in the cosmic
microwave background \cite{Bennett:2003bz}, and the large scale
structure of the universe (e.g., \cite{seljak}) cannot be explained
with a cosmological model that contains only ordinary (baryonic+dark)
matter evolving according to Einstein's theory of general relativity. The most popular
solution is to introduce an extra component with negative
pressure, the so-called dark energy (e.g., \cite{Peebles:2002gy}).

While the simplest candidate for dark energy is a cosmological
constant (or vacuum energy) the extremely small size of the cosmological
constant is difficult to explain unless one is willing to invoke a selection effect (e.g.,
\cite{Weinberg:1988cp}). Furthermore there is an additional strange
coincidence, that the density of photons, neutrinos, baryons and dark
matter and dark energy density all seem to become comparable at roughly the same time in the
history of the universe even though they scale very
differently with redshift.

Dynamical models of dark energy (e.g., quintessence
\cite{quintessence}, k-essence \cite{k-essence,Chiba:1999ka}) may be
built to try to explain the coincidence. These models require an
extremely light scalar field (with mass $\lesssim 10^{-33} {\rm
eV}$). An interesting alternative, which simultaneously alleviates
the need for an extremely light scalar field, and explains the
coincidence between the neutrino and dark energy densities has been
recently proposed in the context of a theory of mass varying
neutrinos (MaVaNs)\cite{kaplan,fardon} (also see \cite{peccei}). In
this model, dark energy is modulated by the density of neutrinos,
which can simultaneously cause the evolution of both dark energy
density and neutrino masses as a function of cosmic time. The time
evolution however is not caused by a slowly rolling light scalar,
but by a scalar (the so-called {\it acceleron}, $\a$), which is
heavy (relative to the Hubble expansion rate) but has a minimum that
varies slowly as a function of the density of neutrinos. The model
may also have interesting implications for neutrino oscillation
experiments \cite{kaplan,Zurek:2004vd} and solar neutrino
observations \cite{Cirelli:2005sg,Barger:2005mn}.

In this work, we investigate the evolution of perturbations in the
MaVaNs model, which has not yet been consistently studied in the
literature. The key difference between perturbations in MaVaNs and
most other dark energy candidates is that (unlike quintessence or
k-essence) all the dynamical properties of (non-relativistic) MaVaNs
are set by the local neutrino density. In particular, pressure is a
local function of neutrino density, implying that the hydrodynamic
perturbations are adiabatic. Subsequently, the speed of sound for
adiabatic perturbations at scales smaller than the Hubble radius is
set by the instantaneous time derivatives of pressure and density
\cite{mukhanov}. We will show that this implies that hydrodynamic
perturbation are unstable on all macroscopic scales once the
neutrinos become non-relativistic.

The paper is organized as follows: in Sec.\ref{MaVaNs}, we summarize
properties of the homogeneous MaVaNs model for dark energy.
Sec.\ref{instability} presents hydrodynamic and kinetic theory
arguments for why perturbations of non-relativistic MaVaNs have an
imaginary sound speed, and are thus {\it unstable}. In particular we
will show that, when neutrinos are non-relativistic, free streaming
is unable to stabilize perturbations on all scales with wavelengths
larger than $m_\a^{-1}$ where $m_\a$ is the mass of the acceleron
field.
We will then move on to speculate about the end result of the
instability.  In Sec.\ref{nuggets}, we argue that the outcome of
instability is inevitably a multi-phase medium with most of
neutrinos in a dense phase, which we call neutrino nuggets, and
dilute as pressureless matter. Sec.\ref{phase} presents a
thermodynamic description of the phase transition and draws the
analogy with the liquid-gas phase equilibrium. Finally,
Sec.\ref{discuss} considers potential stable MaVaNs (or adiabatic
dark energy) models, and Sec.\ref{conclude} concludes the paper. One
thing will become clear: After neutrinos become non-relativistic,
the model has a serious microscopic instability. As a result of this
instability, the fluid will cease to act as dark energy and will not
be able to drive the acceleration of the universe.

\section{Preliminaries of the Dark Energy with Mass-Varying
Neutrinos}\label{MaVaNs}

Within the MaVaNs theory of dark energy, the local energy density
associated with dark energy is the sum of the energy densities of
neutrinos and a light a scalar field, $\a$, the so-called acceleron
field, which also modulates the neutrino mass: \beq V =\nn \mn(\a)+
V_0(\a), \label{v}\eeq where the neutrinos are assumed to be
non-relativistic.

In the limit $\nn \gg m^3_\a$, where $m_\a$ is the mass of the
acceleron field, the acceleron field responds to the average density
of neutrinos, and relaxes at the minimum of the potential $V$ (the
actual relaxation of $\a$ at the minimum, of course, depends on its
evolution through the cosmic history). In this limit,
both $\mn$ and $V_0$ become functions of the neutrino density,
$\nn$, through \beq {\partial V \over
\partial \a}
= \left(\nn + {\partial V_0 \over \partial \mn}\right){\partial \mn
\over \partial \a} =0 \Rightarrow \nn = -{\partial V_0 \over
\partial \mn},\label{nnu}\eeq
if $\partial \mn / \partial \a \neq 0$.

Assuming that the kinetic energy in the acceleron field is
negligible \footnote{One can see that the ratio of acceleron kinetic
to potential energy is $\sim(1+w)^2(H/m_{\a})^2 \ll 1$}, the
equation of state for the neutrino/acceleron fluid is: \beq
 w \equiv \frac{\rm Pressure}{\rm Density} \simeq
 -\frac{V_{0}(\a)}{V} = -1 + {\nn \mn(\a) \over V},\label{w}
 \eeq
where we used the fact that non-relativistic neutrinos have
negligible pressure. We note that $w \simeq -1$ (i.e. a cosmological
constant), as long as the energy density in neutrinos is small
enough compared to the energy of the acceleron field.

Fardon et al. \cite{fardon} go on to propose a specific model for
the mass varying neutrinos: \bea \mn(\a) = {m^2_{lr}\over {\cal
M}(\a)},\label{mna}\\ V_{0}(\a) = \Lambda^4 \ln\left[1+{\cal
M}(\a)/\mu\right] + C , \label{v0a}\eea where it is assumed that
${\cal M}(\a)/\mu \gg 1$, and $\Lambda \sim 10^{-3} {\rm eV}$
characterizes the energy scale of dark energy and $C$ is just a
constant needed to set the minimum of the potential to zero (i.e. no
cosmological constant). Now, using Eq.(\ref{nnu}), we find: \beq V_0
\simeq \Lambda^4 \ln\left(m_0 \over \mn\right) + C \Rightarrow \nn =
\frac{\Lambda^4}{\mn},\label{nlambda} \eeq where $m_0 \equiv
m^2_{lr}/\mu$, and Eq.(\ref{w}) gives the equation of state: \beq w
= -1 + \left[ 1+\ln\left(m_0 \over \mn\right)\right]^{-1}. \eeq

Based on current observational constraints $|1+w| \lesssim 0.2$
(e.g., \cite{seljak}), and thus $ m_0 \gtrsim 10^2 \mn$. Therefore,
we restrict the rest of our analysis (with the exception of Sec.
\ref{phase}) to the $m_0 \gg \mn$ regime.

For the model to be sensitive to the mean density of neutrinos,
there needs to be many neutrinos within the Compton wavelength of
$\a$, i.e. $N_\nu\sim \nn/m^3_\a \gg 1$. The mass of $\a$ is given
by, \beq m_\a^2=\left({\partial^2 V \over \partial A^2}\right
)_{min}= {\Lambda^4 \over \mn^2} \left({\partial \mn \over
\partial \a}\right)^2. \eeq We thus have \beq N_\nu\sim {\mn^2\over
\Lambda^2} \left({\partial \mn \over \partial \a}\right)^{-3}.\eeq
Observational bounds restrict the coupling $({\partial \mn /
\partial \a}) < 10^{-6}$ \cite{fardon} which puts a lower bound on
$N_\nu$ around $10^{24}$, justifying the assumption that $\a$ field
is only sensitive to to the mean neutrino number density.

For non-vanishing $\partial V/ \partial \mn$, Eq.(\ref{nnu}) can
also be satisfied if $\partial \mn /
\partial \a = 0 $ for a value of $\a$ or $\mn$. This can happen
 if $V_0(\a)$ reaches a minimum, which
subsequently puts a maximum on $\mn < m_{max}$. $m_{max}$
corresponds to the mass of neutrinos in vacuum or when $\nn <
\Lambda^4/m_{max}$. In particular, the macroscopic dynamics of the
original MaVaNs model \cite{fardon} is quantified in terms of
$\Lambda$, $m_0$ and $m_{max}$, where \beq V_0(\mn)= \Lambda^4
\ln\left(1+{m_0\over \mn}\right) + C, ~ \mn < m_{max}.
\label{logpot}\eeq Again, $C =-\Lambda^4\ln(1+m_0/m_{max})$ is added
such that the minimum of the potential is at zero.

\section{Emergence of an Imaginary Speed of
Sound}\label{instability}
\subsection{Hydrodynamic picture}\label{hydro}
In the conventional quintessence models \cite{quintessence}, similar
to inflationary models, the scalar field is slowly rolling at the
present epoch, and therefore its effective mass is smaller than the
Hubble expansion rate. In contrast, the acceleron field sits at the
instantaneous minimum of its potential, and the cosmic expansion
only modulates this minimum through changes in the local neutrino
density. The mass of the acceleron field can be much larger than the
Hubble expansion rate. However, by the same token, the coherence
length of acceleron, $m^{-1}_{\a}$, is much smaller than the present
Hubble length, and thus, unlike quintessence, the perturbations on
sub-Hubble scales $> m^{-1}_\a$ are adiabatic, i.e. obey the same
equation of state as the homogeneous universe. Therefore, the speed
of sound, $c^2_{s}$, for these perturbations is simply given by
\cite{mukhanov}: \beq c^2_{s} = \frac{\dot{P}}{\dot{\rho}}=
\frac{\dot{w}\rho+w\dot{\rho}}{\dot{\rho}} =
w-\frac{\dot{w}}{3H(1+w)} = \frac{\partial \ln \mn}{\partial \ln
\nn},\label{cs2w}\eeq where we used the continuity equation
$\dot{\rho} = -3H(1+w)\rho$, and in the last step, we used
Eq.(\ref{nnu}).

We can already see that for a vanishing or small enough $\dot{w}$,
$c^2_s$ has the same sign as $w$, which is negative. Clearly, this
is a catastrophic instability in this theory, as the sub-Hubble
perturbations on scales larger than $m^{-1}_{\a}$ ($H<k<m_{\a}$)
would grow as $\exp(k|c_s|t)$ if $c^2_s <0$.

In particular, for the model introduced in \cite{fardon},
Eq.(\ref{nlambda}) gives: \beq \rho+P = \mn\nn = \Lambda^4 = {\rm
const.} \Rightarrow c^2_s = \frac{\dot{P}}{\dot{\rho}}= -1.\eeq

\subsection{Kinetic theory picture}\label{kinetic}

One may wonder if, similar to the gravitational instability of
collisionless cold dark matter (or massive neutrinos with constant
mass), our instability is stabilized at small scales due to free
streaming of neutrinos. In this section we study the nature of this
instability in the kinetic theory picture, and show that the
imaginary speed of sound is indeed scale-independent in any
non-relativistic MaVaNs models, making the instability most severe
at the microscopic scales ($\sim m^{-1}_{\a}$).

 The action of a test particle with a dynamical mass $\mn({\bf x},\eta)$, in a
flat FRW universe, is given by: \beq S = -\int \mn({\bf x},\eta) ds
= -\int \mn({\bf x},\eta)a(\eta)\sqrt{1-u^2} d\eta, \eeq where \beq
u^2 = {\bf u\cdot u} = {d{\bf x}\over d\eta}\cdot{d{\bf x}\over
d\eta}, \eeq  and $a$ and $\eta$ are the scale factor and conformal
time respectively. Thus, the single particle Lagrangian, Momentum,
and Energy take
the standard form \bea {\cal L}_{\nu} = -\mn a\sqrt{1-u^2}, \\
{\bf p} = {\partial {\cal L}_{\nu} \over
\partial {\bf u}} = \frac{\mn a {\bf u}}{\sqrt{1-u^2}} = \mn\gamma a {\bf u},
\\ {\cal E}_{\nu} = {\bf u}\cdot{\bf p} - {\cal L}_{\nu} = \mn \gamma a,\label{energy}\\ {\rm where~~} \gamma \equiv (1-u^2)^{-1/2},  \eea
while we have dropped the ${\bf x},\eta$ dependence for
brevity.

The evolution of ${\bf p}$ is given by the Euler-Lagrange equation:
\beq {\bf p}^{\prime} = {d{\bf p}\over d\eta} = {\partial {\cal
L}_{\nu} \over
\partial {\bf x}} = - a\gamma^{-1} {\bf \nabla}\mn,\eeq where ${\bf
\nabla}$ denotes the gradient with respect to the comoving
coordinates ${\bf x}$, and we have neglected gravity.

Now the Boltzmann equation for the evolution of the phase space density
of neutrinos, $f({\bf x},{\bf p})$, takes the form: \beq
\frac{\partial f}{\partial \eta} + {\bf u}\cdot {\bf \nabla} f - a
\gamma^{-1}{\bf \nabla}\mn\cdot {\partial f \over
\partial {\bf p}} = 0. \label{Boltzmann}\eeq

Let us also generalize Eq. (\ref{nnu}) for relativistic neutrinos.
Using the fact that neutrino momentum remains constant in a
homogeneous background, the relaxation of the $\a$ field at the
minimum of its potential leads to \beq \frac{\partial V}{\partial
\a} = \left[\int \frac{\mn}{\sqrt{p^2+\mn^2}}f({\bf x,p}) d^3{\bf
p}+\frac{\partial V_0}{\partial \mn}\right]\frac{\partial
\mn}{\partial \a} = 0, \label{dvda}\eeq which for $\partial
\mn/\partial \a \neq 0$ ($\mn < m_{max}$) yields \beq
 \nn \langle \gamma^{-1} \rangle =
 -\frac{\partial V_0}{\partial\mn} \simeq {\Lambda^4\over
 \mn},\label{mngamma}
 \eeq
for the Logarithmic potential.

 In the absence of perturbations, neutrinos become
non-relativistic as they are free-streaming. Therefore, as a result
of momentum conservation, they are frozen into the {\it
relativistic} Fermi-Dirac phase space density: \beq \bar{f}({\bf p})
=\frac{2(2\pi)^{-3}}{\exp\left[|{\bf p}|/T_{0,\nu}\right]+1} =
\frac{2(2\pi)^{-3}}{\exp\left[p_{ph}/T_{\nu}\right]+1},
\label{fermi-dirac}\eeq where $p_{ph} = |{\bf p}|/a$ is the physical
momentum, while $T_{\nu} = T_{0,\nu}/a$ is the neutrino kinematic
temperature, which matches the thermodynamic temperature in the
relativistic regime. The factor of $2$ in Eq. (\ref{fermi-dirac})
accounts for both neutrinos and anti-neutrinos.

Let us study the linear perturbations of the Boltzmann equation (Eq.
\ref{Boltzmann}), in the sub-Hubble regime, i.e. for time/length
scales much shorter than the Hubble time/length. We will show that the instability occurs on microscopic scales so this is perfectly adequate. In this regime,
without loss of generality, we can assume $a = 1$. For plane-wave
linear perturbations: \bea \delta f = \Delta({\bf p})
\exp\left[i({\bf k \cdot x} - \omega \eta)\right], \\ \delta \mn =
\Sigma \exp\left[i({\bf k \cdot x} - \omega \eta)\right], \eea
Eq.(\ref{Boltzmann}) gives: \beq \omega \Delta({\bf p}) = ({\bf
k}\cdot{\bf u}) \Delta({\bf p}) - \gamma^{-1} \left({\bf
k}\cdot\frac{\partial \bar{f}}{\partial {\bf p}}\right)\Sigma,
\label{omegadelta}\eeq while $\partial V/\partial \a =0$ constraint
(Eq. \ref{dvda}) yields \beq \int \gamma^{-1}\Delta({\bf p}) d^3p +
\lambda \Sigma = 0,\label{lambdasigma}\eeq where \bea \lambda =
{\partial^2 V_0\over\partial \mn^2} + \int
\frac{p^2}{(p^2+\mn^2)^{3/2}} \bar{f}(p) d^3p\nonumber\\ =
{\partial^2 V_0\over\partial \mn^2} + \left(\nn\over
\mn\right)\left[{\langle p^2 \rangle\over \mn^2} + O\left(p \over
\mn\right)^4\right]. \label{lambda}\eea

We first notice that that there is no preferred scale in
Eq.(\ref{omegadelta}), as only the combination $\omega/k$ ($= c_s$;
speed of sound), appears in the equation. This is contrary to what happens for gravity where there is a scale, the Jean's scale, below which random motions stabilize perturbations. The reason for the difference is that for scales $k < m_\a$ the acceleron field adjusts to the local density rather than satisfy a Poisson like equation.

Taking the first term on
the right hand side to the left, dividing by $\omega - {\bf k \cdot
u}$, and substituting from Eq. (\ref{lambdasigma}) yields: \beq
\Delta({\bf p}) = \frac{\left({\bf \hat{k}}\cdot \frac{\partial
\bar{f}}{\partial {\bf p}}\right)\int \gamma^{-1}(p^{\prime})
\Delta({\bf p}^{\prime}) d^3{\bf p}}{\gamma(p) (c_s - {\bf \hat{k}
\cdot u}) \lambda}.\eeq We can eliminate the integral of the unknown
amplitude $\Delta({\bf p})$ through multiplying both sides by
$\gamma^{-1}(p)$, and integrating over $d^3{\bf p}$, which give sthe
characteristic equation for $c_{s}$ \beq \lambda = \int {d^3 p \over
\gamma^2(p) (c_s - {\bf \hat{k} \cdot u})}\left({\bf \hat{k}}\cdot
\frac{\partial \bar{f}}{\partial {\bf p}}\right). \label{kdfp}\eeq

In the limit of $u^2 \ll |c^2_s|, 1$, we can expand the argument of
the integral in Eq.(\ref{kdfp}), in powers of $u$, and ignore
$O(u^4)$ terms. The angular parts of the integrals can be taken
using the spherical symmetry of $\bar{f}({\bf p})$, which yields:
\bea \lambda = \int (4\pi p^2 dp) \left(\frac{p}{3\mn
c^2_s}\right)\left(\frac{\partial \bar{f}(p)}{\partial
p}\right)\times\nonumber\\\left[1+\left({3\over 5 c^2_s} -{3\over
2}\right)\left(p\over\mn\right)^2 + O\left(p\over
\mn\right)^4\right] \nonumber\\ = -\frac{\nn}{\mn c^2_s}
\left[1+(c^{-2}_s-5/2){\langle p^2\rangle \over \mn^2}+O\left(p\over
\mn\right)^4\right], \label{lambda2}\eea where, in the last step, we
used
integration by parts.
Combining Eqs. (\ref{lambda}) and (\ref{lambda2}) yields: \beq
c^2_s= -{\nn\over\mn\left(\partial^2 V_0 \over \partial
\mn^2\right)}\left[1+ (c^2_s+c^{-2}_s-5/2){\langle p^2\rangle \over
\mn^2}+O\left(p\over \mn\right)^4\right]\label{c2s1}\eeq

Finally, following Eq.(\ref{mngamma}), the relation between $\nn$ and $\mn$ is also corrected at
finite temperature:
 \beq \nn = -\left[1+{\langle p^2\rangle\over
2\mn^2}+O\left({p\over\mn}\right)^4\right]{\partial V_0 \over
\partial \mn},\eeq which can be substituted in Eq. (\ref{c2s1}) to
give:\bea c^2_s= \mn^{-1}{\partial V_0 \over
\partial \mn} \left(\partial^2 V_0 \over \partial
\mn^2\right)^{-1}\times\nonumber\\ \left[1+
(c^2_s+c^{-2}_s-2){\langle p^2\rangle \over \mn^2}+O\left(p\over
\mn\right)^4\right].\label{c2s2}\eea
%
%
Note that in the limit of $p \ll \mn$, this result is equivalent to
Eq. (\ref{cs2w}) for non-relativistic neutrinos, as $\nn \simeq
-\partial V_0/\partial \mn$ in this limit, so that
\beq \mn^{-1}{\partial V_0 \over
\partial \mn} \left(\partial^2 V_0 \over \partial
\mn^2\right)^{-1} = \frac{\partial \ln \mn}{\partial \ln \nn}
=-1,\eeq where the last equality is for the logarithmic potential.

Moreover, we should point out that the sign of the sound speed is set by the sign of the second derivative of $V(\a)$ at the extremum, $ {\partial^2 V}/{\partial \a^2}\equiv m_\a^2 $. We have  \beq
\frac{\partial^2 V_0}{\partial \mn^2} \simeq \frac{\partial^2
V}{\partial \mn^2} = \frac{\partial^2 V}{\partial \a^2}
\left(\partial \mn \over
\partial \a\right)^{-2}
 {\rm~~for}~~ p^2\ll \mn^2, \eeq
 so that \beq
 c^2_s= {\nn \over \mn m_\a^2 } \left(\partial \mn \over
\partial \a\right)^{2}
\eeq
{\it Therefore, any realistic MaVaNs scenario with $m_\a^2 > 0$
becomes unstable to hydrodynamic perturbations in the
non-relativistic regime}.

 In particular, for the Fermi-Dirac neutrino phase space
density (Eq. \ref{fermi-dirac}), we have $\langle p^2 \rangle
/T^2_{\nu}= 15 \zeta(5)/\zeta(3) \simeq 12.9$, and assuming the
Logarithmic model of Sec. \ref{MaVaNs}, $V_0 \simeq -\ln \mn +{\rm
const.}$, we end up with \beq c^2_s = -1+51.8 \left(T_{\nu} \over
\mn\right)^2+O\left({T_{\nu}\over\mn}\right)^4. \eeq Thus, we see
that for the  logarithmic potential, the model is {\it unstable} if $T_{\nu}
\lesssim \mn/7.2$. Using Eq. (\ref{nlambda}), it is easy to see that
the unstable regime corresponds to $T_{\nu} \lesssim 1.2 \Lambda$.

\section{The Outcome of the Instability: Neutrino
Nuggets}\label{nuggets}

As we argued above, there is no preferred macroscopic ($ > m_\a^{-1}$) scale in the
linear problem.  In particular, the free-streaming of neutrinos acts
uniformly on all macroscopic scales, and may only stabilize the perturbations of
relativistic neutrinos. Therefore, the inevitable outcome of the
instability is the formation of non-linear structures in neutrino
density or {\it neutrino  nuggets}. As the instability is the
fastest at smallest scales  we expect nuggets to form instantly at scales larger than
$m^{-1}_{\a}$, after the onset of instability. The non-linear
nuggets may then slowly merge to form larger nuggets. However, one
may also envision a scenario in which non-relativistic neutrinos are
continuously heated back up to relativistic temperatures, as a
result of instability, and at the expense of reducing the vacuum
energy of the $\a$ field. We investigate this possibility below:

\subsection{Is it possible to have a homogeneous and marginally stable
phase?}\label{stable}

Assuming that this scenario works, neutrinos would be constantly
heated up at microscopic scales through scattering off non-linear
neutrino nuggets. As the instability, and hence scattering occurs at
microscopic scales, neutrinos will have a microscopic mean free
path, which implies that evolution should be adiabatic on
macroscopic scales. The assumption of marginal stability then
implies: \beq c^2_s = \frac{\dot{P}}{\dot{\rho}} = 0 \Rightarrow P =
{\rm const},\eeq which is necessary to maintain homogeneity and
marginal instability at the same time. However, using energy
conservation, $\dot{\rho} + 3H (\rho +P)=0$, we can see that $\rho$
and $P$ can only have a ${\rm\Lambda}$CDM-like behavior: \bea \rho =
A + {B\over a^3}, \\ P = - A, \eea where $A$ and $B$ are integration
constants. In particular, we require that \beq \rho+P = \nn\mn
\langle \gamma(1+{u^2\over 3}) \rangle = {B \over a^{3}},\eeq where
we used the fact that $P = \nn\mn\langle \gamma u^2\rangle/3$ for
the neutrino gas. However, using the fact that $\nn \propto a^{-3}$,
we find \beq \mn \langle \gamma(1+{u^2\over 3}) \rangle = {B\over
n_{0,\nu}} = {\rm const.}\Rightarrow \mn \leq {B\over n_{0,\nu}}.
\eeq The fact that $\mn$ reaches a maximum, implies that either
$V_0(\mn)$ (or $V_0(\a)$) reaches a minimum, or $\langle \gamma^{-1}
\rangle \propto \nn^{-1} \propto a^3$ (eq. \ref{mngamma}). The latter, however, can only
last for less than a Hubble time, as $\gamma \sim 1$ at the onset of
instability, while $\langle \gamma^{-1} \rangle <1$.

Therefore, we see that the only way to maintain homogeneity of
marginally stable neutrino gas is for the acceleron to settle at its
true minimum, following which the neutrinos are decoupled and become
non-relativistic, while the evolution follows the standard
${\rm\Lambda}$CDM scenario.

\subsection{Production of Neutrino Nuggets: Qualitative description of a multi-phase
outcome}\label{qual}

In Sec. \ref{stable} we argued that the instability cannot be
followed by a homogeneous (single phase) state of the
neutrino-acceleron fluid, unless the acceleron settles at its true
minimum (i.e. vacuum). If this does not happen, the outcome can only
be a multi-phase medium.

As result of the instability, the neutrino fluid should fragment
into condensed neutrino nuggets (or the liquid /interactive phase),
within which, neutrinos maintain relativistic energies to have a
stable density distribution. The remaining neutrinos remain in a
tenuous gas phase, and should remain relativistic to maintain their
stability. Moreover during the instability there is a possibility of converting part of the energy into $\a$ particles. The $\a$ particles are not trapped inside the nuggets and their contribution to the energy density diminishes with the Hubble expansion.

As neutrinos in the ``gas phase" continuously lose energy to Hubble
expansion, and nuggets, due to their small volume fraction, cannot
significantly affect the gas phase, most of the neutrinos should
accrete into the nuggets. In fact, as we show in Sec. \ref{phase},
within the assumption of thermal equilibrium, this process instantly
exhausts the neutrinos outside the nuggets, unless they can decouple
(i.e., $\a$ settles at its true vacuum) within the gas phase.

The fraction of neutrinos in the gas phase is set by the equilibrium
of evaporation rate from and accretion rate onto the surface of the
nuggets. However, the large density contrast between inside and
outside of the nuggets implies that only neutrinos with a large
Lorentz factor ($\gamma > m_{out}/m_{in} \simeq n_{in}/n_{out}$) can
escape the nuggets. As $\langle\gamma\rangle$ inside nuggets is
limited by a few times its initial value at the onset of the
instability (where $\langle\gamma\rangle$ was  $ \sim 1$), the
equilibrium can only be maintained when a small fraction of
neutrinos remain outside the nuggets.

Thus we conclude that the most probable outcome of the instability
is the formation of dense non-linear structures of neutrinos,
surrounded by practically empty space where the acceleron has
settled to its minimum. All this is happening at microscopic scales.
As far as the expansion of the universe is concerned, the fluid
energy density will redshift as matter and thus cannot drive the
acceleration of the Universe.

\subsection{Production of $\a$ particles} 

Another possible outcome of the instability is production of $\a$
particles. In fact, one may speculate that, since the characteristic
time and length scale for the instability of non-relativistic MaVaNs
is $m^{-1}_{\a}$, a significant fraction of neutrino/acceleron
energy may end up in $\a$ particles (or accelerons). However, we
should note that, at the onset of the instability, the neutrinos are
still marginally relativistic, implying that the characteristic time
for the instability may be significantly longer.

  Since the speed of sound (and thus the instability) is modulated
  by the Hubble expansion, it is reasonable to approximate $c^2_s$
  as:
  \beq
  c^2_s = - \alpha (H \Delta t) + O(H\Delta t)^2,
  \eeq
   close to the onset of the instability, where $\alpha \sim 1$, is constant
   of order unity. The maximum growth occurs at the scale of $m^{-1}_{\a}$, and is given by
   \beq
   \Delta \propto \exp(|c_s|m_{\a} \Delta t) = \exp(\alpha H^{1/2} \Delta t^{3/2}
   m_\a).
   \eeq
   Therefore, the characteristic growth time for the instability is
   given by
   \beq
   \Delta t_{ins} \sim \left(m_\a \over H\right)^{1/3} m^{-1}_\a \gg
   m^{-1}_\a,
   \eeq
   since, by construction,  acceleron is much heavier than Hubble scale in MaVaNs
   theories (see Sec. \ref{MaVaNs}). This implies that the production of
   $\a$ particles cannot be a significant energy sink, at least during
   the initial stage of
   instability (or nugget formation).

   Subsequent mergers of
   relativistic $m^{-1}_\a$-sized nuggets may potentially lead to $\a$ particle
   production. However, the above argument can also be used to show that
   the velocities of produced nuggets, $v_{nug}$, can never be relativistic:
   \beq
   v_{nug} \lesssim { m^{-1}_\a \over \Delta t_{ins}} \ll 1.
   \eeq

\section{Nugget-Gas Phase Transition: The Thermodynamic
Picture}\label{phase}

In Sec. \ref{qual} we presented a qualitative argument for why most
of neutrinos must condense into nuggets as they become
non-relativistic. In this section, we present a more rigorous
picture of the associated phase transition within the assumption of
thermodynamic equilibrium.

However, we should first note that, in the homogeneous phase, the
thermodynamic approximation is clearly invalid, as the cross-section
for neutrino interaction is small, and thus,  by construction,
thermal relaxation takes much longer than the Hubble time
\cite{fardon}. However, in the relativistic regime, the distribution
remains thermal as the expansion does not change the original shape
of the thermal Fermi-Dirac distribution. After neutrinos become
non-relativistic, the instability develops within microscopic time
and length scales, and thus, it seems plausible to assume that the
system reaches some sort of  thermal equilibrium within microscopic
times.

Assuming thermodynamic equilibrium, the phase space density of
neutrinos is given by the Fermi-Dirac distribution: \beq d\nn =
f(\en) {d^3p \over (2\pi)^3} = 2\left[\exp\left(\en(p)-\mu \over
T\right) +1\right]^{-1} {d^3p \over (2\pi)^3}, \eeq where the
additional factor of $2$ accounts for both neutrinos and
anti-neutrinos. $\mu$ is the chemical potential, which can be
non-zero if the neutrino number is conserved, while $T$ stands for
the thermodynamic temperature.

Similar to previous sections, we assume that the acceleron field,
$\a$, is relaxed at its minimum. Therefore, the system is subject to
the constraint condition of Eq. (\ref{dvda}), which in conjunction
with the condition: \beq \nn = \int {d^3p \over (2\pi)^3} f({\bf
p}), \eeq set the equation of state of the system, as they fix all
thermodynamic variables (e.g., $\mu$ and $\mn$) for given neutrino
number density, $\nn$, and temperature, $T$. In particular, the
pressure of the system is given by: \beq P = \frac{1}{3} \nn \mn
\langle \gamma u^2 \rangle - V_0(\mn). \eeq

\begin{figure}
\includegraphics[width=\linewidth]{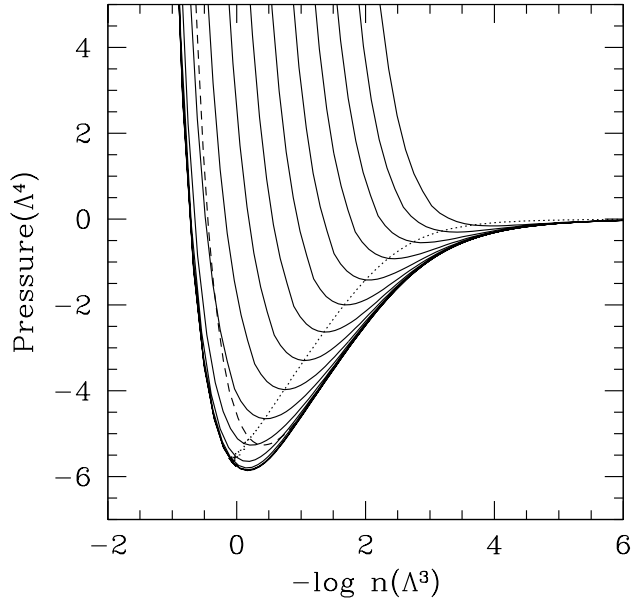}
\caption{\label{pv0} Pressure-Volume phase diagram for the
Logaithmic MaVaNs model with $m_0 = 10^3 \Lambda$ and $m_{max}
\rightarrow \infty$. The solid curves are isothermal contours with
$T = 2^{\ell} \Lambda$, for integer $\ell$'s, where the highest
temperature is $T = 1024 \Lambda$. The isothermal contours become
degenerate below $T \sim 0.5 \Lambda$. The thermal average of
neutrino Lorentz factor, $\langle \gamma \rangle$, is equal to $2$
on the dotted curve, separating the relativistic and
non-relativistic regimes. The dashed curve is the adiabat with $\mu
\rightarrow 0$ as $T \rightarrow \infty$, which approximates the
cosmic history of neutrinos in the absence of inhomogeneities.}
\end{figure}
\begin{figure}
\includegraphics[width=\linewidth]{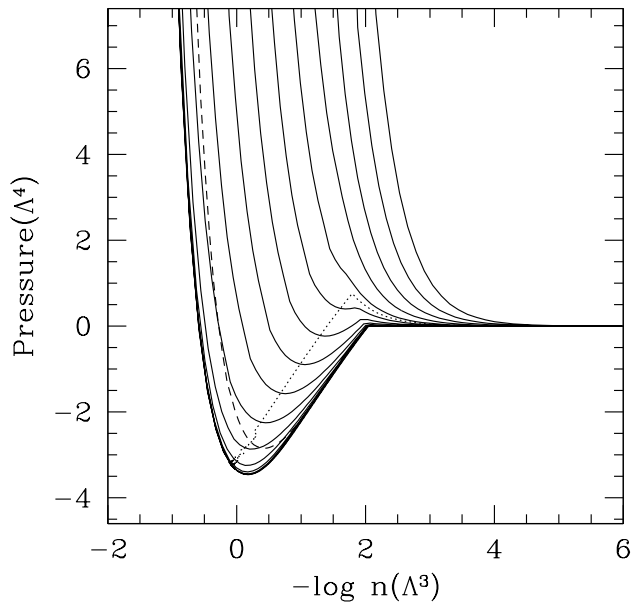}
\caption{\label{pv1} The same as Fig. (\ref{pv0}) for $m_{max} =
10^2 \Lambda$.}
\end{figure}

Fig. (\ref{pv0}) shows the pressure as a function of specific volume
($\equiv \nn^{-1}$) for the system described by these equations for
the logarithmic potential of Sec. \ref{MaVaNs} with $m_0 = 10^3
\Lambda$ (Eq. \ref{logpot}). The solid curves show different
temperatures of $\tn = 2^{\ell} \ \Lambda$ for integer $\ell$'s, where
the highest plotted temperature is $1024\  \Lambda$. The curves become
degenerate below $\tn \sim 0.5\ \Lambda$, as the degeneracy pressure
dominates the thermal pressure. The dotted curve corresponds to
$\langle \gamma \rangle =2$, which separates the relativistic and
non-relativistic regimes.

A homogeneous medium becomes unstable if its pressure decreases with
increasing density. In Sec. \ref{kinetic}, we saw that this happened
when neutrinos become non-relativistic, which is also clearly seen
in Fig. (\ref{pv0}), where $\partial P/\partial n$ changes sign
(i.e. isothermal curves reach a minimum) at the boundary of
relativistic and non-relativistic regimes.

The dashed curve in Fig. (\ref{pv0}) is the adiabat with $\mu=0$ as
$T \rightarrow \infty$, which approximates the cosmic history of
MaVaNs for the logarithmic potential. We again see that the pressure
reaches a minimum (rendering the system unstable), when the
neutrinos become non-relativistic, which also happens close to where
the adiabat crosses $T=\Lambda$ isothermal curve, as demonstrated in
Sec. \ref{kinetic}. Therefore, as the density of neutrinos decreases
monotonically due to cosmic expansion, they become unstable and
fragment into nuggets at $\nn \sim \Lambda^3$, as no stable phase is
available at lower densities. This can also be seen in the
asymptotic (large $\nn$) limit, as the kinetic pressure is simply
given by $\nn T$, which is subdominant to the acceleron pressure for
$\mn \gg m_0$, as $-V_0 =-\Lambda^4 \ln (1+m_0/\mn) \propto
-\nn^{1/2}$, using the mass-density relation in the relevant range.
Therefore, $\partial P / \partial \nn$ remains negative at smaller
densities, and so the system does not have any other stable phase,
justifying the qualitative picture of Sec. \ref{nuggets}.

A more interesting case of a finite maximum neutrino mass, with
$m_{max} = 10^2 \Lambda$ is shown in Fig. (\ref{pv1}). The curves
are the same as in Fig. (\ref{pv0}). Here, we see that for densities
lower than a critical density (which depends on temperature), the
neutrino mass is fixed at its maximum, implying that the acceleron
field relaxes at its true vacuum. Therefore, we end up with a gas of
free-streaming neutrinos with the acceleron settled at the minimum. As a result, the system
shows two stable phases of interactive (or liquid) mass-varying
neutrinos and non-interactive (or gas) neutrinos with $\mn=m_{max}$.
The two phases can co-exist only if they have the same pressure,
temperature and chemical potential.

\begin{figure}
\includegraphics[width=\linewidth]{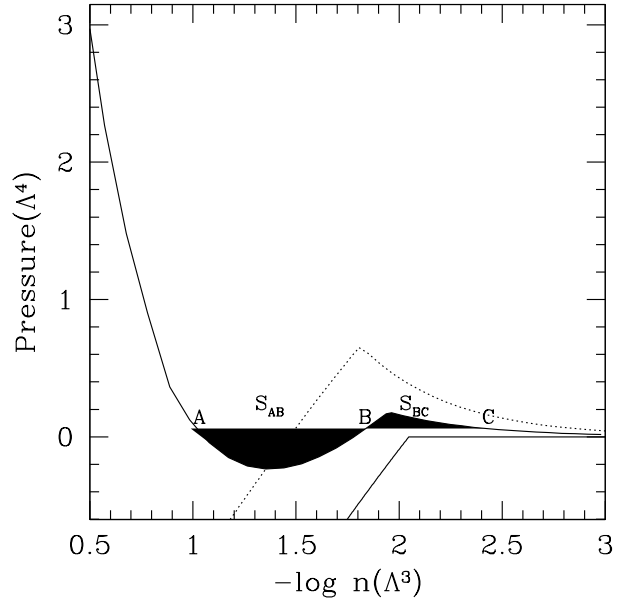}
\caption{\label{pv2} A blow-up of Fig. (\ref{pv1}), which shows the
Gibbs construction for $T = 16 \Lambda$. Points A and C are
interactive and non-interactive (liquid and gas) phases, which can
coexist as they have the same pressure, temperature and chemical
potential. This requires areas $S_{AB}$ and $S_{BC}$ to be equal
(after including the logarithmic metric in the plot), as explained
in the text. The dotted line corresponds to $\langle\gamma\rangle
=2$, while the lower solid line shows the $T=0$ degenerate curve.}
\end{figure}

Fig. (\ref{pv2}) is a blow-up of the $T=16\Lambda$ isothermal
contour from Fig. (\ref{pv1}). Points A and C on the isothermal
contour are liquid and gas phases which can coexist at this
temperature. A pictorial way of finding the coexistence states is
via Gibbs construction. Introducing $\epsilon$ and $s$, as the
energy and entropy per unit particle, the thermodynamic and
extensivity relations subsequently take the form \bea d\epsilon = T
ds -P dn^{-1}, \\ \epsilon = Ts -Pn^{-1} +\mu, \eea which can be
combined to give \cite{kubo} \beq d\mu = n^{-1} dP -s dT. \eeq
Integrating over the isothermal contour ABC, the condition of equal
chemical potentials for the coexisting phases A and C takes the
form: \beq \mu_C -\mu_A = \int_{\rm ABC} n^{-1} dP = S_{AB} - S_{BC}
=0. \eeq Therefore, the area enclosed in regions $S_{AB}$ and
$S_{BC}$ must be the same (even though it may not look this way due
to the logarithmic scale in Fig. (\ref{pv2}).

\subsection{Onset of phase transition: bubble nucleation vs. linear instability}

Even though the neutrino-acceleron fluid does not become unstable to
linear perturbations until it becomes non-relativistic, it can
become unstable to perturbations of finite amplitude, otherwise
known as {\it bubble nucleation}, much earlier. We will argue however that
this is not relevant in the cosmological context being studied.

Consider a bubble of true vacuum with size larger than $m^{-1}_{\a}$
in the relativistic interactive (liquid) phase. If the pressure of
the ambient medium is smaller than the vacuum pressure inside the
bubble (which is set to zero in our treatment), the bubble rapidly
expands until it is in pressure equilibrium with the ambient medium.
This can in principle be achieved through contraction of the medium
as a result of bubble nucleation.

However, the onset of bubble nucleation depends on the statistics of
seed bubbles present in the system. In particular, the smaller the
value of $m_{\a}$, the larger will be the minimum size of the bubble
that can trigger a phase transition. The bounds on $m_{\a}$, which
come from the requirement of no thermal acceleron production in
early universe \cite{fardon}, imply more that $10^{24}$ neutrinos in
a volume of $m^{-3}_{\a}$, making it extremely unlikely for seed
(vacuum) bubbles to exist in a Hubble volume. Cosmological
fluctuations of neutrino fluid are also damped as a result of
free-streaming, and thus cannot seed bubble nucleation. Therefore,
the phase transition is unlikely to start until the pressure reaches
its minimum, where small linear perturbations can seed the
instability.

The outcome of the instability, which develops on microscopic time
scales, can be uniquely obtained from conservation of total energy
and particle densities. For example, for the case of $m_{max}
\rightarrow \infty$ where pressure and density vanish in the vacuum,
the final density of fragmented nuggets is roughly at the
intersection of $\mu=0$ adiabat (dashed curve in Fig. \ref{pv0})
with the $P=0$ axis, which is approximately where $T_{\nu} \sim
\Lambda [\ln(m_0/\Lambda)]^{1/4}$.

\subsection{The fate of nugget-gas multi-phase state}

Finally, let us consider the fate of the system of coexisting
neutrino nuggets and gas. Assuming that $m_0 \gtrsim m_{max} \gtrsim
\Lambda$\footnote{Note that $m_0 \gtrsim m_{max}$ assumption is not
fundamental, and is only made to simplify the expressions. The $m_0
\lesssim m_{max}$ regime can be treated in exactly the same way
which results in a similar behavior to the one obtained in the
text.}, the area of the region $S_{AB}$ (see Fig. \ref{pv2}) is
dominated by the non-relativistic regime and is given by: \beq
S_{AB} \sim \int_{AB} n^{-1} dP \simeq m_{max} +
O\left(\Lambda,{m^2_{max}\over m_0}\right). \eeq

Neglecting the degeneracy pressure (which can be easily justified
for $m_{max} \gtrsim \Lambda$), the area of region $S_{BC}$ is given
by:\beq S_{BC} = \int^{\Lambda^4m^{-1}_{max}}_{n_{gas}} n^{-1} T dn
= -T \ln(n_{gas} m_{max}/\Lambda^4). \eeq Therefore, the Gibbs
construction $S_{AB} = S_{BC}$ yields \bea n_{gas} m_{max} =
\Lambda^4 \exp\left(-\frac{m_{max}}{T}\right) \ll \Lambda^4
\nonumber \\{\rm ~for~} T \lesssim \Lambda \lesssim
m_{max}.\label{gibbs}\eea As we anticipated in Sec. \ref{qual}, as
long as the phase transition occurs far enough from the true minimum
of $V_0(\mn)$, the density of neutrinos in the gas phase (where $\a$
relaxes at its true vacuum) is exponentially suppressed. The
subsequent evolution follows from requiring energy conservation:
\bea d ({\bar n}_{\nu} a^3 T) =  -{\bar n}_{\nu} a^3 m_{max}
d\left[1\over \ln(n_{gas} m_{max}/\Lambda^4)\right] \nonumber\\=
-n_{gas} T da^3 = {n_{gas} m_{max} da^3 \over \ln(n_{gas}
m_{max}/\Lambda^4)}, \eea which can be approximately solved to give:
\bea \frac{n_{gas}}{{\bar n}_{\nu}} \propto {T \over m_{max}} =
-{1\over \ln(n_{gas} m_{max}/\Lambda^4)} \nonumber\\\sim {1\over
m_{max}/\Lambda + 3 \ln (a/a_{\Lambda})} ,\eea where ${\bar
n}_{\nu}$ denotes neutrino cosmic mean density, and $a_{\Lambda}$ is
the cosmological scale factor at phase transition (where $T\sim
\Lambda$). Therefore, both the fraction of neutrinos in the gas
phase, as well as the equilibrium temperature, remain essentially
unchanged their cosmic history.

\section{Discussion}\label{discuss}

\subsection{How about a light neutrino
mass eigenstate?}\label{light}

We have shown that unless the acceleron field sets the mass of a relativistic particle, hydrodynamic perturbations are unstable. The Big Bang model predicts a kinematic temperature of $T_{\nu}
\simeq 0.7 \ T_{\rm CMB} \simeq 1.7 \times 10^{-4}\ev$ for relic
cosmological neutrinos (e.g., \cite{paddy}).  Current experimental
limits on neutrino oscillations require the most massive neutrino
eigenstate to have $m_{\nu} \gtrsim 10^{-2} \ev$ \cite{particle}.
Therefore, at least one of the neutrino eigenstates for cosmological
neutrinos is non-relativistic.


One may imagine a scenario where the interaction
with $\a$ is suppressed for the non-relativistic (including the most
massive) eignestates, and thus the acceleron/dark energy dynamics is
modulated by the density of a relativistic neutrino eigenstate. Even this seems difficult.
For the logarithmic potential, the stability of this
scenario requires $T_{\nu} \gtrsim 1.2 \Lambda$ (Sec.
\ref{kinetic}), which is already constrained, up to a logarithimc
factor, by the current dark energy density ($\Omega_{_{DE}} \simeq
0.7$): \bea \left(2.3 \times 10^{-3} \ev\right)^4 = \rho_{_{DE}}
\simeq \Lambda^4 \ln\left(m_0/\mn\right) \nonumber\\\lesssim 0.5
\ \ T^4_{\nu} \ln\left(m_0/\mn\right) \simeq (1.4 \times 10^{-4}\ev)^4
\ln\left(m_0/\mn\right),\eea requiring \beq\ln(m_0/\mn) \gtrsim 7
\times 10^4,\eeq which does not seem realistic.

Even if we do not restrict
$V_0(\mn)$ to a logarithmic form,
%
assuming a Fermi-Dirac distribution (Eq. \ref{fermi-dirac}), we have
\beq 1+w \simeq \frac{0.6 \ \ T^4_{\nu}}{\rho_{_{DE}}} \sim 10^{-5},\eeq
for any relativistic MaVaNs theory, implying a very slow evolution
that is indistinguishable from the $\Lambda$CDM scenario. One may
hope to be able to test this model by measuring neutrino
oscillations (or structure formation) at high redshifts, as neutrino
mass scales as the inverse of the temperature of neutrino background
in such models ($\mn \propto T^{-1}_{\nu} \propto (1+z)^{-1}$).

\subsection{Adiabatic dark energy and cosmological
constant}\label{adiabatic}

In Sec. \ref{stable} we showed that an adiabatic dark energy model
with $c_s=0$ behaves exactly like a $\Lambda$CDM cosmology. The
argument can be easily generalized for any stable dark energy model
($c^2_s
> 0$) which satisfies the null energy condition ($\rho_{_{DE}}+P_{_{DE}} >0$) in the
following way. Combining the stability condition and conservation of
energy we have \beq
 c^2_s = {\dot{P}_{_{DE}}\over\dot{\rho}_{_{DE}}} = - {\dot{P}_{_{DE}}\over 3H
 (\rho_{_{DE}}+P_{_{DE}})} >0, \eeq
 which can be integrated to give
 \bea
 P_{_{DE}}(t\rightarrow\infty) =\nonumber\\ P_{_{DE}}(t_0) -
 3\int^{\infty}_{t_0} H(t) c^2_{s}(t)\left[\rho_{_{DE}}(t)+P_{_{DE}}(t)\right]
 dt \nonumber\\ < P_{_{DE}}(t_0) < 0.\label{pde}\eea

Since dark energy has negative pressure at the present time
($P_{_{DE}}(t_0)<0$), Eq. (\ref{pde}) guarantees that the pressure
remains {\it negative and finite}, even when all particle densities
approach zero at future infinity. Therefore a stable adiabatic dark
energy model will necessarily asymptote to a cosmological
constant/vacuum energy at future infinity (or it goes unstable).

\section{Conclusions}\label{conclude}

We have demonstrated that the model of dark energy with mass-varying
neutrinos (MaVaNs)\cite{fardon,kaplan} is subject to a
linear instability with an imaginary speed of sound. The onset of
instability is around the time that neutrinos become
non-relativistic, and is due to the fact that, unlike quintessence
perturbations, the fluctuations in the non-relativistic neutrino
density evolve adiabatically. The acceleron field mediates a strong attractive force (stronger than gravity) between neutrinos. There are more than $10^{24}$ neutrinos inside the Compton wavelength of the $\a$ field. When neutrinos become non-relativistic, their random motions are not sufficient to stop  collapse and the medium becomes unstable.

The question of the outcome of the instability, an inherently
non-linear process, is a more difficult problem to address. We
present qualitative arguments for why the outcome of the instability
should be a multi-phase medium, where most of neutrinos end up in
dense nuggets (interactive/liquid phase) with nearly constant
density, while a small fraction may remain outside in a decoupled
gas phase. We then confirm these arguments quantitatively within the
assumption of thermodynamic equilibrium, which is plausible
following the phase transition. Both nuggets, and free neutrinos
(which are exponentially suppressed) dilute similar to cold dark
matter particles, with negligible pressure, leaving a vacuum
energy/cosmological constant as the only possible source of the
observed cosmic acceleration after the onset of instability in the
MaVaNs models.

\acknowledgments It is our pleasure to thank Nima Arkani-Hamed for
valuable discussions. This work is partially supported by NSF grants
AST 0098606 and by the David and Lucille Packard Foundation
Fellowship for Science and Engineering and by the Sloan Foundation.


\bibliography{catast}

\end{document}